\documentclass[preprint]{mn2e}
\usepackage{color}
\usepackage{graphicx}
\usepackage{dcolumn}
\usepackage{bm}

\begin{document}

\newcommand{\myspace}{1.0}
\renewcommand{\baselinestretch}{\myspace}
\Large\normalsize

\title[Triaxial density--potential--force profiles]
{Realistic triaxial density--potential--force profiles for stellar systems
and dark matter halos}

\author[Terzi\'c \& Sprague]
{Bal{\v s}a Terzi\'c$^{1,2}$ and Benjamin J. Sprague$^1$\\ 
$^1$NICADD, Department of Physics, Northern Illinois University,  DeKalb, IL 60115, USA\\
$^2$Corresponding Author: bterzic@nicadd.niu.edu}

\date{Received 2007 January 4}

\pubyear{2007} \volume{000} 
\pagerange{\pageref{firstpage}--\pageref{lastpage}}

\input{psfig}

\maketitle
\label{firstpage}

\begin{abstract}
Popular models for describing the luminosity-density profiles of dynamically 
hot stellar systems (e.g., Jaffe, Hernquist, Dehnen) were constructed to match 
the deprojected form of de Vaucouleurs' $R^{1/4}$ light-profile.  However, 
we now know that elliptical galaxies and bulges display a mass-dependent
range of structural profiles.  To compensate this, the model in 
Terzi\'c \& Graham was designed to closely match the deprojected form of 
S\'ersic $R^{1/n}$ light-profiles, including deprojected exponential 
light-profiles and galaxies with partially depleted cores.  It is thus 
applicable for describing bulges in spiral galaxies, dwarf elliptical 
galaxies, both ``power-law'' and ``core'' elliptical galaxies, also dark 
matter halos formed from $\Lambda$CDM cosmological simulations.
In this paper, we present a new family of triaxial density-potential-force 
triplets, which generalizes the spherical model reported in Terzi\'c \& 
Graham to three dimensions.  If the (optional) power-law core is present, 
it is a $5$-parameter family, while in the absence of the core it 
reduces to $3$ parameters.  The isodensity contours in the new family are 
stratified on confocal ellipsoids and the potential and forces are 
expressed in terms of integrals which are easy to evaluate numerically.  
We provide the community with a suite of numerical routines for orbit 
integration, which feature: optimized computations of potential and 
forces for this family; the ability to run simulations on parallel platforms;
and modular and easily editable design.
\end{abstract}


\section{Introduction}
Elliptical galaxies, bulges of spiral galaxies and dark matter halos 
exhibit a range of density-profile shapes which are accurately described by 
the S\'ersic (1963, 1968) $R^{1/n}$ model (e.g. Caon, Capaccioli \& 
D'Onofrio 1993; Young \& Currie 1994; Graham et al.~1996; Graham 2001;
Balcells et al.~2003; Merritt et al.~2006).  
This model is a generalization of de Vaucouleurs' 
(1948, 1959) $R^{1/4}$ model which provides a decent fit only to galaxies 
with $M_B \approx -21$ mag (e.g.  Kormendy \& Djorgovski 1989; 
Graham \& Guzm\'an 2003).  The popularity of de Vaucouleurs' model permeated 
into computer modeling of elliptical galaxies: the most popular models 
used in galaxy dynamics simulations -- Jaffe (1983), Hernquist (1990) and 
Dehnen (1993, see also Tremaine et al. 1994) -- were designed to approximate 
the $R^{1/4}$ light profile when projected.  All three of these density 
models are double power-law models and have the same fixed outer profile 
slope, declining with radius as $r^{-4}$.  This means that, while 
undeniably useful, these models are limited in their ability to model 
realistic galactic structures and their evolution (Terzi\'c \& Graham 2005).

More sophisticated power-law models have since been developed to allow for 
more flexibility in matching deprojected galaxy light profiles.  
A 3-parameter Dehnen double power-law model was generalized by Zhao (1996) 
into a 5-parameter model such that both the inner and outer power-law slope 
can be adjusted, along with the radius, density and sharpness of the 
transition region.  This model was first introduced by Hernquist 
(1990, his equation 43) and has the same structural form as the ``Nuker'' 
model (e.g., Lauer et al.\ 1995).  However, Graham et al.\ (2003) show that 
such double power-law models do not provide an adequate description of 
profiles with logarithmic curvature, i.e.\, profiles without inner and outer 
power-laws which is the case for the luminosity-density profiles of most 
elliptical galaxies and bulges.  The 3-parameter model of Prugniel \& Simien 
(1997) was, however, designed to describe profiles with curvature and does 
therefore not suffer from such a problem.  Their density model provides a 
good analytic approximation to the deprojected S\'ersic $R^{1/n}$ model.

These expressions have proven useful for describing the density
profiles of dark matter halos simulated in cosmological models of
hierarchical structure formation (Merritt et al.\ 2006; Graham et al.\ 2006).
Merritt used techniques from nonparametric function 
estimation theory to extract the density profiles, and their derivatives, 
from a set of $N$-body halos, to show that the $3$-parameter Prugniel--Simien 
density profile provides a better description to the data than other 
$3$-parameter models currently in use, such as the generalized NFW 
(Navarro, Frenk, \& White 1996) model with arbitrary inner slope.

Recently, Terzi\'c \& Graham (2005) developed a 5-parameter model which is 
able to unite properly the domain of the black hole with the outer parts of 
the galaxy.  It modifies the 3-parameter Prugniel--Simien density model 
to allow for (optional) partially depleted power-law cores, providing 
excellent match to the deprojected light profiles of elliptical
galaxies, including the ones with depleted cores for which the 
Prugniel--Simien model is inadequate.  Expressions for the potential and 
forces corresponding to the spherical Prugniel--Simien 
model are reported in Terzi\'c \& Graham (2005).

Double power-law models owe a good deal of their popularity to the fact 
that the expressions for the potential and forces are relatively simple, 
in some cases even expressible in terms of analytic functions.  
This leads to easy implementation of computer routines for orbital 
integration.  In the trade-off between faithfulness to the physical 
problem (as measured by how accurately these density profiles match the 
observed deprojected light-profiles of elliptical galaxies) on one side, and 
numerical simplicity and computational efficiency on the other, these 
models favor the latter.  However, after more than two decades of rapid 
progress in computer technology since the first of these models was devised, 
this choice becomes more difficult to justify.  Here we propose to bias the 
trade-off in the opposite direction -- we devise a more realistic physical 
model at the cost of making it mathematically more intricate and 
computationally more demanding.  Our rationale is that if we are to construct 
realistic models of galaxies, it is imperative that we use a model capable 
of reflecting, as accurately as possible, what is observed in Nature.  
We then estimate computational efficiency of this realistic model and 
compare it to that of traditional power-law models.  Finally, we design 
and make freely available an optimized suite of computer programs used for 
orbit integration, in hope to eliminate numerical implementation as an 
adverse factor.

\begin{figure*}
\begin{center}
\includegraphics[width=6.5in]{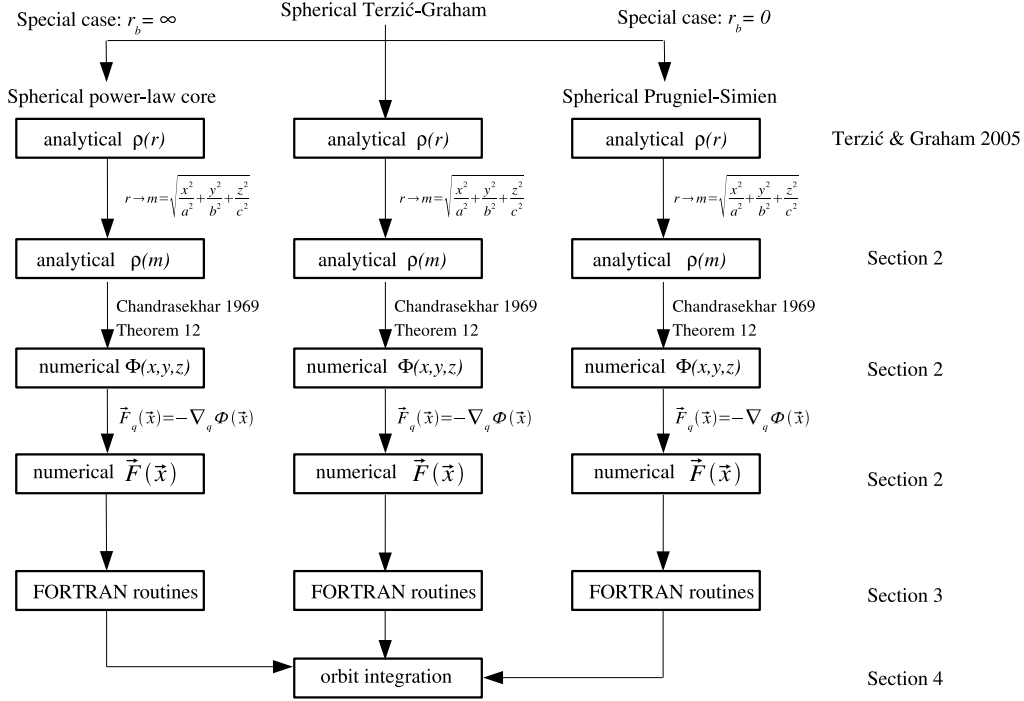}
\vskip-25pt
\caption{Flow-chart outlining the work presented here.}
\label{fig1}
\end{center}
\end{figure*}

In this paper we generalize spherical profiles of Terzi\'c \& Graham (2005)
to three dimensions by replacing radial with ellipsoidal symmetry in the
mass density.  The outline of the new triaxial model in the context of our
earlier work is shown in Fig.~\ref{fig1}.  
In Section \ref{Section:Model}, the new triaxial model is presented: we 
start with ellipsoidally stratified mass density distribution and derive 
the corresponding potential and forces, including special cases when the 
5-parameter model reduces to a 3-parameter deprojected S\'ersic profile 
with no power-law core in Section \ref{ss:PS97} and 2-parameter power-law 
in Section \ref{ss:plc}.  The potential and forces are derived in terms 
of integrals , which are easy to evaluate numerically. 
In Section \ref{Section:Code}, we outline the main features of the numerical 
suite used for orbital integration in the new potential, including code
optimization and parallelization, user interface and information analysis. 
In Section \ref{Section:Comparison}, we use the new model to describe the
bulge at the center of the Milky Way as well as a pair of galaxy profiles.  
We also compare its speed and the goodness of fit to the popular 
Dehnen model \cite{D93,Tetal94}. 
Finally, we summarize and discuss the importance of the model presented 
here in Section \ref{Section:Summary}. 

\section{Triaxial Model} \label{Section:Model} 

In this Section we introduce a family of expressions associated with the 
spatial density profiles of triaxial stellar systems having S\'ersic-like 
profiles with optional power-law cores, generalizing our earlier work on 
equivalent spherically symmetric profiles (Terzi\'c \& Graham 2005).
We derive {\it exact} expressions for the potential and forces associated 
with this family of spatial density profiles in terms of quadratures.  
This profile subsumes two other profiles as its special cases: 
the 3-parameter Prugniel--Simien profile in the limit of the break radius 
$r_b \to 0$, and the 2-parameter power-law core in the limit 
of $r_b \to \infty$.

\subsection{Density} \label{ss:Density} 

The triaxial mass density model is a triaxial generalization to the 
spherical model of Terzi\'c \& Graham (2005) model.  Replacing radial 
dependence with a dependence on an ellipsoidal coordinate $m$, defined as
\begin{equation} \label{m}
m \equiv \sqrt{{{x^2}\over{a^2}} + {{y^2}\over{b^2}} + {{z^2}\over{c^2}}},
\end{equation}
yields a profile with equidensity surfaces stratified on confocal ellipsoids
with axis ratios $a:b:c$
\begin{equation} \label{density}
\rho(m) =
\left\{
  \begin{array}{ll}
\rho_b \left({r_b\over m}\right)^{\gamma}
& \mbox{if $m \le r_b$} \\
\rho_b {\bar \rho}
\left({m\over R_{\rm e}}\right)^{-p} {\rm e}^{ -b_n\left(m/R_{\rm e}\right)^{1/n}}
& \mbox{if $m > r_b$}
  \end{array} \right.
\end{equation}
with
\begin{equation} \label{rhobar}
{\bar \rho}=\left({r_b\over R_{\rm e}}\right)^{p}
{\rm e}^{b_n\left(r_b / R_{\rm e}\right)^{1/n}}.
\end{equation}
Parameters above are defined in the context of the spherical model and its
fits to the observed light profiles \cite{TG05}.
The break radius $r_b$ marks the position of instantaneous transition from
one regime to another.  $\rho_b$ is the density at the break radius.
$\gamma$ is the inner power-law slope, not to be confused with incomplete
gamma function $\gamma[a,x]$ defined later in the text.
$R_{\rm e}$ is the (projected) effective half-light radius when $r_b=0$.  
The parameter $n$ is the S\'ersic index and describes the curvature of 
the profile.  The term $b_n$ is not a parameter, but instead a function 
of $n$ and chosen to ensure $R_{\rm e}$ contains half the (projected) 
galaxy light.  It is obtained by solving the equation 
$\Gamma[2n] = 2 \ \gamma[2n,b_n]$, where
\begin{eqnarray} \label{eqgam}
\Gamma[a] & = & \int_0^{\infty}{\rm e}^{-t}t^{a-1} {\rm d}t,
\hskip20pt \\
\gamma[a,x] & = & \int_0^{x}{\rm e}^{-t}t^{a-1} {\rm d}t, \hskip10pt  a>0 \\
\Gamma[a,x] & = & \Gamma[a] - \gamma[a,x] = \int_x^{\infty}{\rm e}^{-t}t^{a-1} 
{\rm d}t, \hskip10pt  a>0
\end{eqnarray}
are the complete, incomplete and complement to the incomplete gamma functions,
respectively.
A good approximation of $b_n$ for $n>0.5$ is $2n-1/3+0.009876/n$ 
(Prugniel \& Simien 1997; see MacArthur, Courteau, \& Holtzman 2003 
for smaller values of $n$).
Similarly to $b_n$, $p$ is also not a parameter, but a function of $n$,
well approximated by $p = 1.0 - 0.6097/n + 0.05563/n^2$, for $0.6<n<10$.

\subsection{Potential and Forces} 
\label{ss:PotForce} 

The potential arising from an ellipsoidally stratified density distribution
$\rho=\rho (m^2)$ can be written as (Chandrasekhar 1969, page 52, Theorem 12):
\begin{equation} \label{chandra}
\Phi(x,y,z)=-\pi a b c G \int\limits^{\infty}_{0}
{{\psi(\infty) - \psi({\bar m}^2)} \over
 {\sqrt{(\tau+a^2)(\tau+b^2)(\tau+c^2)}} } \ d \tau,
\end{equation}
with
\begin{equation} \label{m_bar}
{\bar m}^2 (\tau) = {x^2 \over {a^2 + \tau}} +
                    {y^2 \over {b^2 + \tau}} + {z^2 \over {c^2 + \tau}},
\end{equation}
and
\begin{equation} \label{psi}
\psi(m^2)=\int\limits^{m^2}_{1} \rho  ({\tilde m}^{2}) \ d {\tilde m}^{2}.
\end{equation}

After some algebra, we find for $\gamma \ne 2$:
\begin{eqnarray} \label{potential_1}
\Phi(x,y,z) & = & 
{{\alpha}\over{2-\gamma}} r_b^{\gamma} H(x,y,z;a^2,b^2,c^2,2-\gamma) 
\nonumber \\
& + & \alpha K R(x,y,z;a^2,b^2,c^2), 
\end{eqnarray}
when $m \le r_b$, and
\begin{eqnarray} \label{potential_2}
\Phi(x,y,z) & = &
{{\alpha}\over{2-\gamma}} r_b^{\gamma} H^I(x,y,z;a^2,b^2,c^2,2-\gamma,\tau_b) 
\nonumber \\
&+& \alpha K R^I(x,y,z;a^2,b^2,c^2,\tau_b) \nonumber \\
&-& \alpha n {\bar \rho} R_e^2 b_n^{n(p-2)} T(x,y,z;a^2,b^2,c^2,\tau_b),
\end{eqnarray}
when $m > r_b$, where
\begin{equation} \label{alpha}
\alpha = 2\pi a b c G \rho_b 
\end{equation}
\begin{equation} \label{K}
K = -{{1}\over{2-\gamma}} r_b^2 - n {\bar \rho} R_e^2 b_n^{n(p-2)} 
\Gamma\left[n(2-p),b_n\left({{r_b}\over{R_e}}\right)^{1/n}\right],
\end{equation} 
\begin{equation} \label{spec_fun_HI}
H^I(x,y,z;A,B,C,q,s) \equiv \int\limits_s^{\infty} 
{{{\bar m}^q ~ d \tau}\over{d_1(\tau;A,B,C)}},
\end{equation} 
\begin{equation} \label{spec_fun_H}
H(x,y,z;A,B,C,q) \equiv H^I(x,y,z;A,B,C,q,0), 
\end{equation} 
\begin{equation} \label{spec_fun_RI}
R^I(A,B,C,s) \equiv \int\limits_s^{\infty} 
{{d\tau}\over{d_1(\tau;A,B,C)}}, 
\end{equation} 
\begin{equation} \label{spec_fun_R}
R(A,B,C) \equiv R^I(A,B,C,0), \\
\end{equation} 
\begin{equation} \label{spec_fun_T}
T(x,y,z;A,B,C,s) \equiv \int\limits_0^s 
{{\Gamma\left[n(2-p),b_n\left({{{\bar m}}\over{R_e}}\right)^{1/n}\right] d\tau}
\over{d_1(\tau;A,B,C)}}, 
\end{equation} 
and
\begin{equation} \label{d1}
d_1(\tau;A,B,C) \equiv \sqrt{(\tau+A)(\tau+B)(\tau+C)}.
\end{equation}
Superscript $I$ denotes an improper integral with a finite non-zero lower 
limit and infinity as an upper limit.  $R(A,B,C)$ is known as Carlson integral
\cite{C88}.  $\tau_b$ is found by solving a cubic equation 
$r_b = {\bar m}(\tau_b)$.

For $\gamma=2$, the expressions for potential are:
\begin{eqnarray} \label{potential_1ex}
\Phi(x,y,z) & = &
\alpha r_b^2 H_l(x,y,z;a^2,b^2,c^2)  \nonumber \\
&+& \alpha K_2 R(a^2,b^2,c^2),
\end{eqnarray}
when $m \le r_b$, and
\begin{eqnarray} \label{potential_2ex}
\Phi(x,y,z) & = &
\alpha r_b^2 H_l^I(x,y,z;a^2,b^2,c^2,\tau_b) \nonumber \\
&+& \alpha K_2 R^I(a^2,b^2,c^2,\tau_b) \nonumber \\
&-& \alpha 
n {\bar \rho} R_e^2 b_n^{n(p-2)} T(x,y,z;a^2,b^2,c^2,\tau_b),
\end{eqnarray}
when $m > r_b$, where
\begin{equation} \label{K2}
K_2 = - r_b^2 \log r_b - n {\bar \rho} R_e^2 b_n^{n(p-2)} 
\Gamma\left[n(2-p),b_n\left({{r_b}\over{R_e}}\right)^{1/n}\right],
\end{equation} 
and
\begin{equation} \label{spec_fun_HlI}
H_l^I(x,y,z;A,B,C,s) \equiv \int\limits_s^{\infty} 
{{\log {\bar m} ~ d\tau}\over{d_1(\tau;A,B,C)}}, 
\end{equation} 
\begin{equation} \label{spec_fun_Hl}
H_l(x,y,z;A,B,C) \equiv H_l^I(x,y,z;A,B,C,0).
\end{equation} 

The gravitational forces for $\gamma\ne 2$ are given by
\begin{equation} \label{forces_1x}
F_{q_i}(x,y,z) = -\alpha r_b^{\gamma} q_i H_2(x,y,z;k_i^2,a^2,b^2,c^2,-\gamma),
\end{equation}
when $m \le r_b$ and
\begin{eqnarray} \label{forces_2x}
F_{q_i}(x,y,z) =
& - &\alpha r_b^{\gamma} q_i H^I_2(x,y,z;k_i^2,a^2,b^2,c^2,-\gamma,\tau_b) 
\nonumber \\
& - & \alpha {\bar \rho} q_i S(x,y,z;k_i^2,a^2,b^2,c^2,\tau_b),
\end{eqnarray}
when $m > r_b$.  $k_1=a$, $k_2=b$, $k_3=c$ and $q_1=x$, $q_2=y$, $q_3=z$.
The quadratures in the above expressions are defined as
\begin{equation} \label{spec_fun_HI2}
H^I_2(x,y,z;D,A,B,C,q,s) \equiv \int\limits_s^{\infty} 
{{{\bar m}^q ~  d\tau}\over{d_2(\tau;D,A,B,C)}}, 
\end{equation} 
\begin{equation} \label{spec_fun_H2}
H_2(x,y,z;D,A,B,C,q) \equiv H_2^I(x,y,z;D,A,B,C,q,0),
\end{equation} 
\begin{equation} \label{spec_fun_S}
S(x,y,z;D,A,B,C,s) \equiv \int\limits_0^s 
{{\left({{\bar m}\over{R_e}}\right)^{-p} 
e^{-b_n\left({{{\bar m}}\over{R_e}}\right)^{1/n}}d\tau}
\over{d_2(\tau;D,A,B,C)}},
\end{equation} 
where
\begin{equation} \label{d2}
d_2(\tau;D,A,B,C) \equiv (\tau+D) ~ d_1(\tau;A,B,C). 
\end{equation}

For $\gamma=2$, no special care in derivation of forces is needed, so they 
coincide with equations (\ref{forces_1x}) and (\ref{forces_2x}).

The expressions (\ref{spec_fun_HI}), (\ref{spec_fun_RI}), (\ref{spec_fun_HlI}), 
(\ref{spec_fun_HI2}) and (\ref{spec_fun_T}), as well as (\ref{spec_fun_S}) when 
$s \to \infty$, are improper integrals and are not very suitable for numerical 
computation.  This problem is resolved by introducing a new variable of 
integration $\xi = (1+\tau)^{-1/2}$.  
The integrals then become proper and are given by
 
\begin{equation} \label{spec_fun_HIt}
H^I(x,y,z;A,B,C,q,s) = 
2 \int\limits_0^{{1\over{\sqrt{1+s}}}} 
{{{\bar m}^{q} ~ d\xi}
\over{{\tilde d}_1(\xi;A,B,C)}},
\end{equation} 
\begin{equation} \label{spec_fun_RIt}
R^I(A,B,C,s) = 
2 \int\limits_0^{{1\over{\sqrt{1+s}}}} 
{{d\xi}\over{{\tilde d}_1(\xi;A,B,C)}},
\end{equation} 
\begin{equation} \label{spec_fun_Tt}
T(x,y,z;A,B,C,s) = 
2 \int\limits_{{1\over{\sqrt{1+s}}}}^1 {{
{\Gamma\left[n(2-p),b_n\left({{{\bar m}}\over{R_e}}\right)^{1/n}\right]} d\xi}
\over{{\tilde d}_1(\xi;A,B,C)}},
\end{equation} 
\begin{equation} \label{spec_fun_HlIt}
H_l^I(x,y,z;A,B,C,s) = 
2 \int\limits_0^{{1\over{\sqrt{1+s}}}} 
{{{\log \bar m}^{q} ~ d\xi}\over
{{\tilde d}_1(\xi;A,B,C)}},
\end{equation} 
\begin{equation} \label{spec_fun_HI2t}
H^I_2(x,y,z;D,A,B,C,q,s) = 
2 \int\limits_0^{{1\over{\sqrt{1+s}}}} {{
{\bar m}^{q} \xi^2 d\xi}\over{{\tilde d}_2(\xi;D,A,B,C)}},
\end{equation} 
\begin{equation} \label{spec_fun_St}
S(x,y,z;D,A,B,C,s) =
2 \int\limits_{{1\over{\sqrt{1+s}}}}^1 {{
\left({{\bar m}\over{R_e}}\right)^{-p} 
e^{-b_n\left({{{\bar m}}\over{R_e}}\right)^{1/n}} \xi^2 d\xi
}\over{{\tilde d}_2(\xi;D,A,B,C)}},
\end{equation} 
where
\begin{equation} \label{mbar}
{\bar m}^2 = \xi^2 \left[{{x^2}\over{1+(a^2-1)\xi^2}} +
{{y^2}\over{1+(b^2-1)\xi^2}} + {{z^2}\over{1+(c^2-1)\xi^2}}\right],
\end{equation}
and
\begin{eqnarray} \label{dtilde1}
& & {\tilde d}_1(\xi;A,B,C) \equiv \nonumber \\
& & \sqrt{(1+(A-1)\xi^2)(1+(B-1)\xi^2)(1+(C-1)\xi^2)}, 
\end{eqnarray}
\begin{equation} \label{dtilde2}
{\tilde d}_2(\xi;D,A,B,C) \equiv 
(1+(D-1)\xi^2) ~ {\tilde d}_1(\xi;A,B,C).
\end{equation}

Total mass of the galaxy represented by our new triaxial model is easily found 
to be
\begin{equation} \label{Mtot_tg}
M_{TG} = 2 \alpha 
\left[ {{r_b^3}\over{3-\gamma}} + {\bar \rho} R_e^3 n b^{n(p-3)}
\Gamma\left[n(3-p),b\left({r_b\over R_e}\right)^{1/n}\right]\right].
\end{equation}
To ``normalize" our model to have the total mass equal to unity, the 
expressions for potential and forces need to be divided by $M_{TG}$.

In the spherical limit, $a=b=c=1$, the expressions of Terzi\'c \& Graham 
(2005) are recovered (see Appendix A). 

\subsection{Special Case $r_b=0$: Triaxial Generalization of Prugniel--Simien 
Profile}
\label{ss:PS97} 

In the limit of the break radius $r_b$ going to zero, our two-regime model 
reduces to a single regime S\'ersic profile.  The corresponding equations for 
mass density, potential and forces for this triaxial profile are readily 
obtained after setting $r_b=0$ in the equations (\ref{density}), 
(\ref{potential_2}) and (\ref{forces_2x}), respectively, and are given by
\begin{equation} \label{density_ps}
\rho (m) = \rho_b {\bar \rho}
\left({m\over R_{\rm e}}\right)^{-p} {\rm e}^{ -b_n\left(m/R_{\rm e}\right)^{1/n}},
\end{equation}
\begin{equation} \label{potential_ps}
\Phi(x,y,z) =
-\alpha n {\bar \rho} R_e^2 b_n^{n(p-2)} T(x,y,z;a^2,b^2,c^2,\infty),
\end{equation}
\begin{equation} \label{forces_ps}
F_{q_i}(x,y,z) = -\alpha {\bar \rho} q_i S(x,y,z;k_i^2,a^2,b^2,c^2,\infty), 
\end{equation}

Total mass for the Prugniel--Simien model is readily obtained by substituting
$r_b=0$ into (\ref{Mtot_tg}):
\begin{equation} \label{Mtot_ps}
M_{PS} = 2 \alpha 
{\bar \rho} R_e^3 n b^{n(p-3)} \Gamma\left[n(3-p)\right].
\end{equation}

\subsection{Special Case $r_b=\infty$: Triaxial Power-Law Core}
\label{ss:plc} 

In the limit of the break radius $r_b$ tending to infinity, our two-regime 
model reduces to a single power-law core profile.  The equations for mass 
density, potential and forces for the 3D power-law core are given by
\begin{equation} \label{density_pl}
\rho(m) =
\rho_b \left({r_b\over m}\right)^{\gamma},
\end{equation}
\begin{equation} \label{potential_pl}
\Phi(x,y,z) =
\alpha {{1}\over{2-\gamma}} r_b^{\gamma} r^{2-\gamma} 
H(x,y,z;a^2,b^2,c^2,2-\gamma),
\end{equation}
\begin{equation} \label{forces_pl}
F_{q_i}(r,\theta,\phi) = -\alpha r_b^{\gamma} q_i H_2(x,y,z;k_i^2,a^2,b^2,c^2,-\gamma),
\end{equation} 
 
For some special cases, such as $\gamma=0,1,2$, the above expressions are 
given in terms of elementary functions and Carlson integrals \cite{ZP88,PM01}.
Single power-law model is scale-free, which means that orbital properties 
simply scale with radius.  This property leads to the separation of radial
and angular dependence in spherical coordinates, thus allowing for efficient
series approximation, such as double Fourier or Fourier-Legendre \cite{T02}.

Total mass for the power-law core is readily found to be 
\begin{equation} \label{Mtot_plc}
M = {{2 \alpha}\over{3-\gamma}} r_{\rm max}^3, 
\end{equation}
where $r_{\rm max}$ is the outer radius of the core.

\subsection{Non-Dimensionalizing the Physical Problem}

In order to make the numerical problem dimensionless, we adopt a convention 
$G = M_{\rm tot} = M_{TG} = M_{PS} = M_D = 1$.  This means that, since  
\begin{eqnarray} \label{G}
G & = & 6.672 \times 10^{-11}[{\rm kg}]^{-1}[{\rm m}]^3[{\rm sec}]^{-2} \nonumber \\
  & = & [10^{11} M_\odot]^{-1}[{\rm kpc}]^3 [1.491 \times 10^6 {\rm yr}]^{-2},
\end{eqnarray}
the unit of time is
\begin{equation} \label{t_units}
1.491 \times 10^6 {\rm yr} ~ \sqrt{{M_\odot ~ 10^{11}}\over M}~
{\left({\beta\over{1 {\rm kpc}}}\right)}^{3\over 2},
\end{equation}
where $\beta$ is the length-scale of the model.  The total stellar mass can 
be easily computed from the best-fit parameters of three models, via equations 
(\ref{Mtot_tg}), (\ref{Mtot_ps}) and (\ref{Mtot_d}).
If the observed effective half-light radius is taken to be the length-scale 
of the model $\beta$ (see Table \ref{Tab_Mod}), respective time-scales are 
easily computed using expression (\ref{t_units}).  Each unit of length 
corresponds to 1 kpc.

\begin{figure*}
\begin{center}
\includegraphics[width=6.5in]{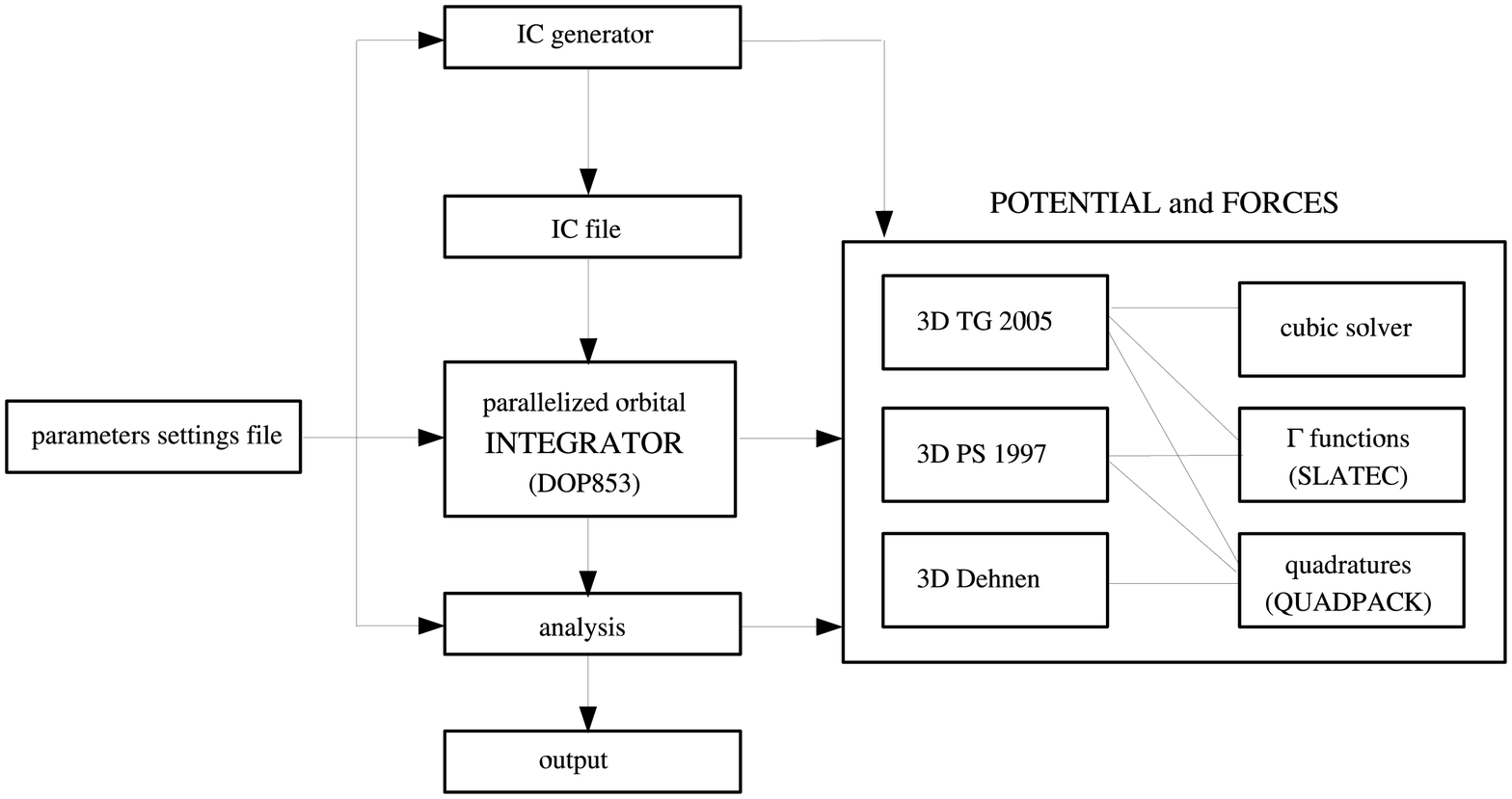}
\caption{Flow-chart outlining the numerical suite.}
\label{figint}
\end{center}
\end{figure*}

\section{Numerical Implementation} \label{Section:Code}
In this Section we outline the major features of the numerical suite which 
we designed to optimize orbit integrations in the new triaxial potential --
triaxial generalization of a 5-parameter Terzi\'c--Graham model, 
as well as its 3-parameter special case (when central core is not depleted)
triaxial Prugniel--Simien model. 

Our goal is to design a state-of-the-art, user-friendly numerical package
which would allow for easy setup and efficient execution of numerical 
integrations of orbits in the new model.  Our suite of numerical routines is
thoroughly tested and fully optimized.  It is also parallelized using MPI, 
with its efficiency scaling roughly linearly with the number of processors 
used.  We make this numerical suite freely available for downloading from our
group's webpage: {\tt http://www.nicadd.niu.edu/research/beams/TS\_2007/}.

\subsection{Outline of the Numerical Suite}
The numerical suite utilizes a number of tested and optimized non-proprietary 
numerical packages written in FORTRAN 77.  The numerical algorithm used for 
integrating orbits is the explicit imbedded (7,8) pair of Dormand and Prince 
to compute orbital points \cite{HNW93}.  Complete and incomplete gamma 
functions are computed by routines from SLATEC, created by American National 
Laboratories (available online at {\tt {http://www.netlib.org/slatec}}).  
Evaluation of quadratures, both with finite and infinite ranges, is done by 
QUADPACK routines \cite{qp}, (also available online at 
{\tt {http://www.netlib.org/quadpack}}).  The driver program is written in 
FORTRAN 90. 

Flow-chart outlining the numerical suite is shown in Fig.~\ref{figint}.  

\subsection{Modular Design}

We designed the integrator core of the numerical package to be as simple as 
possible while allowing easy addition of multiple models and analysis routines 
in a parallelized integrator system.  The data flow of the system is outlined 
in Fig.~\ref{figint}.  A parameter settings file is defined by the user, 
which defines settings for the integrator, the initial conditions generator, 
the potential and forces models, and the analysis routines.  This file is 
first processed by the initial conditions generator to produce an initial 
conditions file.  The integrator core then uses both the parameter settings 
file and the initial conditions file to begin integrating the orbits.  For 
each orbit, the initial conditions are read directly from the initial 
conditions file, and the integrator core obtains the forces from the models 
using the standard ``potential and forces'' interface in order to perform the 
total orbital integration.  When the integration is complete, orbital data
is passed on to the analysis routines, which are responsible for data 
processing and producing the output file for the simulation.

Every effort is made to separate the user-editable code from the system-level 
code.  This allows the system to be modified for use with any model and 
analysis routines without requiring changes to the source files containing the 
integrator or parallelization code.  The user can supply any model to the 
integrator using the standard ``potential and forces'' interface.  This is 
done by simply filling in the supplied functions {\tt get\_acceleration} and 
{\tt get\_potential} with user-specific code.  Supplying analysis routines such 
as an FFT routine or an energy-conservation measurement is similarly trivial, 
by simply adding code to perform the operation in the {\tt analyze\_orbit}
routine.  More advanced analyses such as tracking Lyapunov exponents or 
tangent dynamics are obtained through a similar interface which allows an 
arbitrary number of extra integration variables to be tracked along with the 
main orbit without changes to the main integrator source code.

We created a parallel implementation of this integrator using MPI by 
using a block-decomposition technique to assign nearly equal amounts of 
orbits to each processor.  Each processor computes its orbits one at 
a time, and writes its output to its own local output file.  Once all 
processors have computed their sets of orbits, the main output file is 
constructed by concatenating the individual output files together.  
The final output file contains the orbits in exactly the same order as they 
were arranged in the initial conditions file because a block-decomposition 
assignment of orbits is used.  
The parallel implementation is completely transparent to the user-editable 
code in the analysis and model definition routines, because the parallel 
decomposition of the problem is done in groups of orbits in the same manner 
as the analysis decomposition. 

\subsection{Parallel Implementation}
Our parallel decomposition of the orbital integrator requires very little 
communication between the computation nodes, which is why we estimate that the 
speedup derived from parallel computation scales roughly linearly with the 
number of processors used.  However, this nearly optimal speedup is limited 
by two main factors.  First, the number of orbits to integrate must exceed 
the number of processors.  This is required in order to allow each processor 
to contribute toward the overall solution.  Second, the length of time to 
integrate each orbit must be approximately equal, because the algorithm is 
not finished until the slowest processor has completed its equal sized block 
of orbit integrations.

In common use of this integrator, however, these factors are not expected to 
play a large role in the actual performance of the simulation.  The first 
problem is averted by the fact that integration batches are commonly grouped 
in larger numbers than are typical for a multiple-processor cluster.  
Common clusters have processor counts in the hundreds, while it is often 
desirable to run integration groups numbering from hundreds to thousands of 
stars in each energy level.  The second problem is avoided because it is 
common practice to group initial conditions in batches according to energy 
level.  Since the dynamical time is roughly equal for all stars of the 
same energy level, we can say that all integrations of the same simulation 
time length will take approximately the same amount of real time to simulate.
One exception to this is the fact that centrophilic orbits, ones with no
angular momentum, typically take longer to integrate: the integrator is forced
to take increasingly smaller steps near the very center in order to attain
prescribed accuracy when orbits' velocities are high.  This is why in our 
numerical suite generates initial conditions for centrophilic orbits 
(vanishing angular momentum) separately from the centrophobic ones
(see Appendix B).  Therefore, if these two types of orbits are 
integrated separately, the roughly linear scaling of simulation time with 
the number of processors is preserved.

\section{Model Comparison} \label{Section:Comparison}

The Dehnen model and simple power-law core models have been quite popular as 
{\it qualitative}, ``toy'' models which capture general features of stellar 
dynamics near galactic centers \cite{MF96,VM98,MV99,PM01,T02}.  By devising 
parametric models which provide excellent fits to the full range of the
observed data, not only the central region, we are now in position to develop 
{\it quantitative} models for individual galaxies.

Here we propose the 3-parameter triaxial generalization of Prugniel--Simien 
model as an alternative to the 3-parameter Dehnen model for 
modeling dynamics of elliptical galaxies without depleted cores.  For the
galaxies with depleted cores, we propose a triaxial generalization of the
5-parameter Terzi\'c--Graham.  We first reiterate the findings 
of Terzi\'c \& Graham (2005) that these alternatives provide 
much better fits to deprojected observed light-profiles of realistic 
galaxies using two representative examples from their set of eight elliptical
galaxies and the bulge of the Milky Way.  We then demonstrate that using 
these improved triaxial models in dynamical simulations will not incur 
appreciable additional computational cost over the traditional Dehnen model.  
It is our belief that having a considerably more accurate model of the 
observed physical system more than justifies the modest increase in 
computational expenses.

\begin{figure*}
\begin{center}
\includegraphics[width=6in]{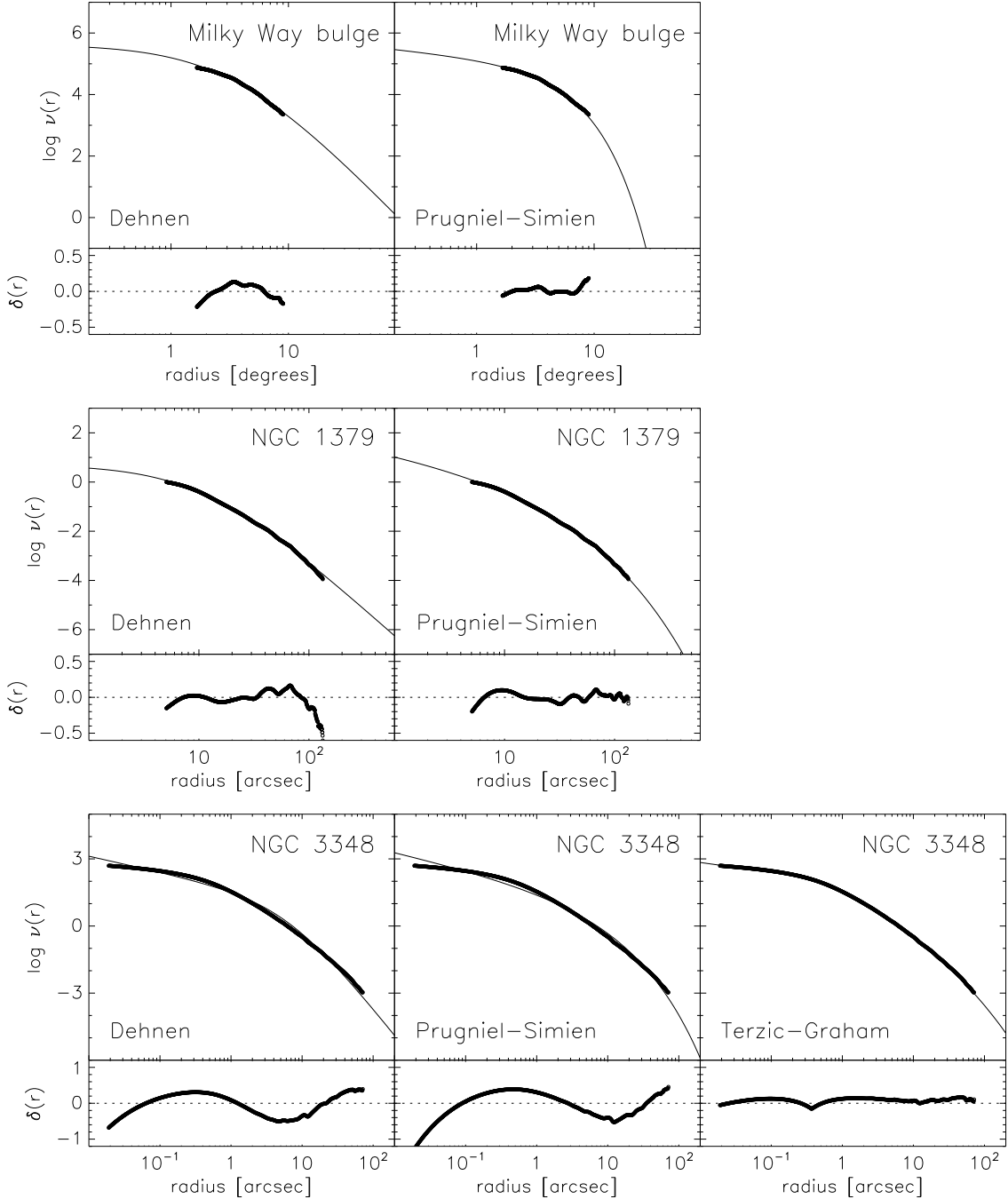}
\vskip25pt
\caption{
Top row: parametric fits (along with the relative error shown underneath) 
of Dehnen model (left), Prugniel--Simien model (right) to the deprojected 
major-axis, $K$-band light profiles of the central bulge in the Milky Way
(fitted out to $R=9$ degrees; data from Kent, Dame \& Fazio 1991).
Middle row: parametric fits (along with the relative error shown 
underneath) of Dehnen model (left), Prugniel--Simien model (right) to the 
deprojected major-axis, $B$-band light profiles of NGC 1379 elliptical galaxy
(data from Caon, Capaccioli \& D'Onofrio 1993).
Bottom row: parametric fits (and relative errors) of Dehnen model (left), 
Prugniel--Simien model (middle) and Terzi\'c--Graham (right) to 
deprojected major-axis, $R$-band light profiles of NGC 3348 elliptical galaxy
(data from Trujillo et al.\ 2004). 
The luminosity-density profile $\nu(r) = \rho(r)/(M/L)$ is in units of 
${\rm L}_{\sun}~{\rm pc}^{-3}$.
The model parameters are given in Table \ref{Tab_Mod}.
}
\label{figd}
\end{center}
\end{figure*}

\subsection{Fitting Realistic Deprojected Light-Profiles}

Terzi\'c \& Graham (2005) fitted several of the most popular density models: 
$2$-parameter Jaffe (1980) and Hernquist (1990); $3$-parameter Dehnen 
and Prugniel--Simien; and their own new $5$-parameter hybrid model to 
deprojected light profiles of several elliptical galaxies.  For their sample 
of luminosity-density profiles taken from real galaxies the $2$-parameter 
Jaffe and Hernquist double power-law models systematically fail to fit the 
entire extent of observed galaxy profiles.  The $3$-parameter Dehnen model
does better, providing a good match to density profiles of large early-type
galaxies with S\'ersic index around $4$ or greater, but it is inadequate for
galaxies with low S\'ersic indices, which include dwarf ellipticals and most
bulges in disk galaxies.  The $3$-parameter Prugniel--Simien model, 
however, accurately matches the density profiles of {\it both} luminous 
elliptical galaxies with $n>4$ and faint elliptical galaxies with $n<4$, 
including exponential ($n=1$) profiles.  For galaxies with partially depleted 
cores, Terzi\'c \& Graham (2005) developed a $5$-parameter, two-regime model, 
capable to match simultaneously {\it both} the nuclear and global stellar 
distribution.  Terzi\'c \& Graham (2005) arrive to these conclusions after 
fitting a set of eight elliptical galaxies of various sizes and light-profile 
shapes.  

In the present study, we focus on the bulge in the center of the Milky Way
and the two galaxies from the sample used in Terzi\'c \& Graham (2005): an 
elliptical galaxy NGC 1379, and an elliptical galaxy with partially depleted 
core -- NGC 3348.  The Milky Way's bulge and the first galaxy, NGC 1379, 
with S\'ersic index of $n=2.0$ (see Table 2 of Terzi\'c \& Graham 2005), 
were chosen because it illustrates that Dehnen is inadequate to describe 
galaxies with low S\'ersic indices, and that Prugniel--Simien model is a 
good alternative.  The second galaxy, NGC 3348, is an example of a galaxy 
with partially depleted cores for which only 5-parameter Terzi\'c--Graham 
model provides good fits in density space.

The profiles have been calibrated in units of ${\rm L}_{\sun}~{\rm pc}^{-3}$ 
using the distance moduli provided in Tonry et al.\ (2001) for NGC 1379 and
the distance of $41$ Mpc for NGC 3348.  We used solar absolute magnitudes
of $M_K=3.33$, $M_B=5.47$ and $M_R=4.28$ (Cox 2000).  A Hubble constant 
$H_0=75 {\rm km}~{\rm s}^{-1}~{\rm Mpc}^{-1}$ is used.

%
In the top two rows of Fig.~\ref{figd}, the $3$-parameter Dehnen 
and Prugniel--Simien models were fitted to the deprojected major-axis
$K$-band light-profile of the Milky Way's bulge (first row) and 
$B$-band light-profile of NGC 1379 (second row).  It is clear that Dehnen 
model is morphologically inadequate to provide a good fit for these types 
of galaxies which have a low value of S\'ersic index $n$.  Prugniel--Simien 
model does better, especially fitting the outer curvature of the profile,
as evidenced by the smaller relative error $\delta(r)$.
In the bottom row of Fig.~\ref{figd}, the $3$-parameter Dehnen 
and Prugniel--Simien models, along with the $5$-parameter
Terzi\'c--Graham were fitted to the deprojected major-axis $R$-band
light-profile of a core elliptical galaxy NGC 3348.  Both Dehnen and
Prugniel--Simien models fail to account for the inner partially
depleted core.  The $5$-parameter Terzi\'c--Graham model does quite
well and is the only density model that can fit galaxies possessing 
depleted cores.

\begin{figure*}
\begin{center}
\includegraphics[width=6.3in]{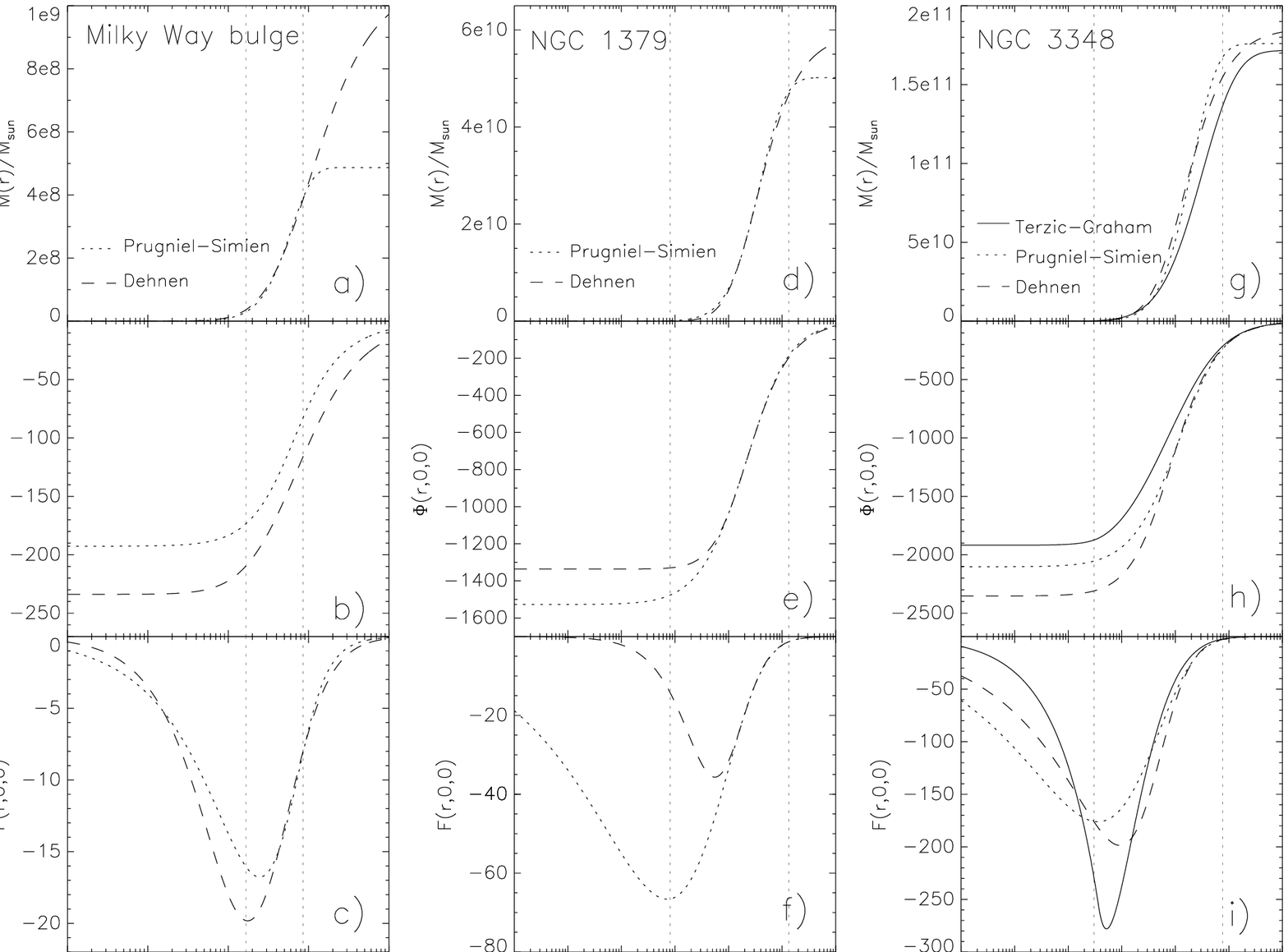}
\vskip30pt
\caption{
Top column: mass enclosed within radius $r$ in terms of $M_{\sun}$
(top), scaled potential (middle) and force (bottom), along the $x$-axis, for
a spherical Dehnen model (dashed lines) and Prugniel--Simien model (dotted
lines) for the Milky Way bulge.
Middle column: mass enclosed within radius $r$ in terms of $M_{\sun}$
(top), scaled potential (middle) and force (bottom), along the $x$-axis, for
a spherical Dehnen model (dashed lines) and Prugniel--Simien model (dotted
lines) for NGC 1379 elliptical galaxy.
Right column: mass enclosed within radius $r$ in terms of $M_{\sun}$ (top),
scaled potential (middle) and force (bottom), along the $x$-axis, for a
spherical Dehnen model (dashed lines), Prugniel--Simien model (dotted
lines) and Terzi\'c--Graham (solid lines) for NGC 3348 elliptical galaxy.
The model parameters are given in Table \ref{Tab_Mod}.
The axis ratios are $1:1:1$.}
\label{figmpf}
\end{center}
\end{figure*}

Fig.~\ref{figmpf} shows the total enclosed stellar mass, scaled 
potential and forces for the models from Fig.~\ref{figd}, using the
best-fitting parameters reported in Table \ref{Tab_Mod}.

\begin{table*}
\caption{
For the bulge at the center of the Milky Way, NGC 1379 and the core 
galaxy NGC 3348: 
total stellar mass, unit of model (simulation) time in physical time 
and the Hubble time ($1.37 \times 10^{10}$ years) in terms of model 
(simulation) time, as computed from the parameters for the three models 
via equation (\ref{t_units}).  We adopt the observed half-light radius 
computed from S\'ersic and core-S\'ersic fits to projected light-profiles, 
reported in Table 2 of Terzi\'c \& Graham (2005), as the length-scale 
$\beta$.  Mass-to-light ratios are taken to be $M/L=0.66$ for the $K$-band, 
$M/L=3$ for the $R$-band, and $M/L=5.3$ for the $B$-band (Worthey 1994).
}
\label{Tab_Hubble}
\begin{tabular}{|lclccc|}
\hline
Gal. Id. & $\beta$ & Model & Total stellar mass & Unit of time & $T_{\rm Hubble}$\\
         &     [arcsec]                &       & $[M_{\sun}]$       & [years]      & [model units]   \\
\hline
Milky Way & $1.65\times 10^4$         &Dehnen&$1.00\times 10^{9}$& $4.60\times 10^{6}$  &  $2.98 \times 10^{3}$ \\
         &                             &Prugniel--Simien &$5.34\times 10^{8}$& $6.30\times 10^{6}$  &  $2.17 \times 10^{3}$ \\
\hline
NGC 1379 & 24.3                        &Dehnen&$5.93\times 10^{10}$& $4.50\times 10^{6}$  &  $3.04 \times 10^{3}$ \\
         &                             &Prugniel--Simien &$5.02\times 10^{11}$& $4.90\times 10^{6}$  &  $2.78\times 10^{3}$ \\
\hline
NGC 3348 & 21.4                        &Dehnen &$1.86\times 10^{11}$& $6.11\times 10^{6}$  &  $2.24\times 10^{3}$ \\
         &                             &Prugniel--Simien&$1.76\times 10^{11}$& $6.29\times 10^{6}$  &  $2.18 \times 10^{3}$ \\
         &                             &Terzi\'c--Graham&$1.72\times 10^{11}$& $6.37\times 10^{6}$  &  $2.15 \times 10^{3}$ \\
\hline
\end{tabular}
\end{table*}

\begin{table*}
\caption{Best-fitting parameters from the three density models over the entire
range of the deprojected profile.  The units are ${\rm L}_{\sun}~ 
{\rm pc}^{-3}$ for $\rho$ and arcseconds for $R_{\rm e}$, $r_a$ and $r_b$.
}
\label{Tab_Mod}
\begin{tabular}{|lcccccccccccc|}
\hline
Gal. Id. & Band & \multicolumn{3}{c}{|------ Dehnen ------|} &
\multicolumn{3}{c}{|-- Prugniel--Simien --|} &
\multicolumn{5}{c|}{|--------- Terzi\'c--Graham ---------|} \\
  &  &  $r_a$  & $\log \rho(r_a)$ & $\gamma_D$ & $R_{\rm e}$  & $\log \rho(R_{\rm e})$  &  $n$  & $R_{\rm e}$ & $n$ & $\gamma$ & $r_b$ & $\log \rho(r_b)$ \\
\hline
Milky Way bulge & $K$ & $1.20\times10^4$ & -7.31 & 0.00 & $1.57\times10^4$ & -7.52 &  1.08 & ...  & ...  &  ... & ... & ... \\
NGC 1379 & $B$ & 11.1 & -0.49 & 0.00 & 24.7 & -1.31 &  2.10 & ...  & ...  &  ... & ... & ... \\
NGC 3348 & $R$ & 6.40  &  0.14 & 0.71 & 13.2 & -0.63 &  2.10 & 20.2 & 3.6 & 0.44 & 0.37 & 2.15 \\
\hline
\end{tabular}
\end{table*}

\subsection{Computational Efficiency} \label{subsection:speed}

The quadratures involved in computing forces for the Dehnen model are similar
to those used for Prugniel--Simien and Terzi\'c--Graham models (compare
the equation (\ref{Dehnen_f}) to equations (\ref{forces_1x})-(\ref{forces_2x}) 
and (\ref{forces_ps})).  It is therefore reasonable to expect that alternative  
implementations of the numerical evaluation of these quadratures will be 
equally reflected on the computational speeds.  In other words, the relative
computational efficiency outlined later in the Section should be general.

It is possible to considerably improve computational efficiency of force
computations by exploiting simplifications arising for special cases of 
parameters (e.g. integer values of the central cusp $\gamma_D$ for the
Dehnen model) or galaxy shapes (e.g. spherical case $a=b=c=1$).  These
simplifications were exploited in the earlier, {\it qualitative} studies
featuring the Dehnen model, in which $\gamma_D=0$ represented the model
with no central cusp, $\gamma_D=1$ ``weak'' and $\gamma_D=2$ ``strong'' 
central cusp (Friedman \& Merritt 1996, Siopis 1999).  However, for 
{\it qualitative} studies such as the one we are proposing here, it is not
expected for deprojected galaxy profiles to be best fitted by parameter
values which take on these fortuitous values, which is why these 
simplifications arising from special cases are not explored here.

\subsection{Comparing Speed of Orbital Integration}
We now compare numerical efficiency of orbital integrations of the two new 
models -- Terzi\'c--Graham and its special case Prugniel--Simien -- with the 
traditional Dehnen model for a typical galaxy.  (Expressions for density, 
potential and forces for the Dehnen model are given in Appendix C.)  
We compare integration times for equivalent sets of orbits in the three 
models obtained after parametric fitting in density space to the deprojected 
light-profile of NGC 3348 (Table \ref{Tab_Mod}).

In comparing the computational speeds of various models against one another, 
it is necessary to determine a setting which provides a fair comparison. 
On the surface, it seems that comparing run times for orbital integrations 
of equal integration time would be a reliable setting.  In fact, this 
is the case when using a constant time-step ordinary differential 
equation (ODE) solver, because it guarantees that the {\tt get\_acceleration} 
routine will be called an identical number of times for each model.  

However, restricting the orbital integrator to constant time-step solvers is 
not desirable because of their poor performance when compared to adaptive-step
solvers.  Adaptive-step solvers adjust the size of the integration step so as 
to satisfy the predefined error tolerance.
Simulation time for integrations using adaptive-step ODE solver only obtains 
physical significance when scaled to {\it dynamical time} of the orbit at a 
certain energy level.  In this setting, we define the dynamical time to be the 
average period of an oscillation of a star around the origin of the galaxy.  

For the NGC 3348 (parameters are given in Table \ref{Tab_Mod}), we integrated 
600 box orbits and 600 tube orbits (see Appendix B) at 50 logarithmically 
spaced energy levels in the range $E(10^{-2}\beta) \le E(r) \le E(5\beta)$ 
(where $E(r)=\Phi(r,0,0)$ denotes a isoenergy surface pierced by the $x$-axis
at $x=r$), for roughly 3 Hubble times (length-scale $\beta$ and Hubble times 
are given in Table \ref{Tab_Hubble}).  We used two axis ratios, that of 
spherical model $a=b=c=1$, and that of a maximally triaxial 
($(a^2-b^2)/(a^2-c^2)=0.5$) $a:b:c=1:0.79:0.5$.

\begin{figure}
\begin{center}
\includegraphics[width=3in]{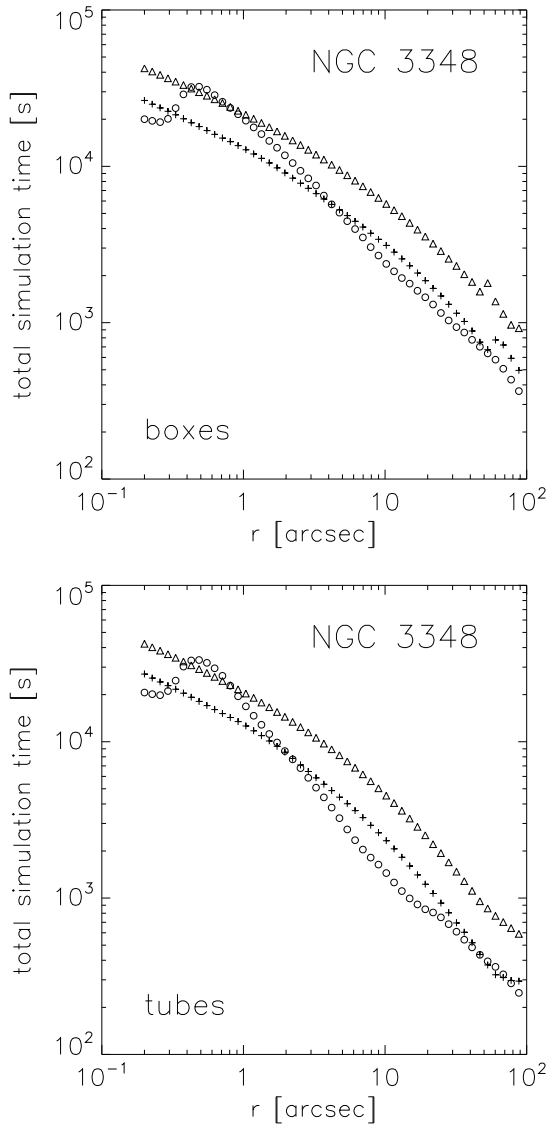}
\vskip15pt
\caption{Total simulation time for 600 orbits at varying energy levels 
(denoted by the maximum radial excursion) in potentials approximating 
NGC 3348 galaxy, integrated for the equivalent of 3 Hubble times.  
Top row: box orbits (left) and tube orbits (right) in the spherical geometry 
(axis ratio $1:1:1$).
Bottom row: box orbits (left) and tube orbits (right) in the maximally
triaxial geometry (axis ratio $1:0.79:0.5$).  
Crosses represent Dehnen model, triangles Prugniel--Simien and 
circles Terzi\'c--Graham.
}
\label{figtime1}
\end{center}
\end{figure}

Figure \ref{figtime1} shows the total simulation time.
The Prugniel--Simien is consistently about $40-80\%$ slower than 
the Dehnen model, while Terzi\'c--Graham, on average, is at least as fast 
as the Dehnen model.  There is a ``bump'' in the line corresponding to the 
Terzi\'c--Graham model around the break radius.  This is due to the fact 
that the adaptive-step ODE integrator samples the region around the transition 
more densely, which results in increase in the total simulation times.  

Increasing the steepness of the negative central density slope 
($\gamma_D$ for Dehnen, $\gamma$ for Terzi\'c--Graham and $p$ for 
Prugniel--Simien) of the model renders orbital integration more 
computationally intensive, because of the rapid changes in the forces.  
For NGC 3348, the central density slopes of the Dehnen 
($\gamma_D=0.71$) is nearly identical to that of the Prugniel--Simien 
($p=0.72$), but somewhat larger than that of Terzi\'c--Graham 
($\gamma=0.44$).  This means that the comparison of execution speeds for 
orbits in potentials modeling NGC 3348 will accurately reflect efficiency of 
force evaluations in the Prugniel--Simien model relative to that of the 
Dehnen model, while the Terzi\'c--Graham will have a slight ``advantage'', 
having the shallower central slope.  This is indeed what we see in 
Fig.~\ref{figtime1}: the 5-parameter Terzi\'c--Graham model outperforms 
both 3-parameter models -- the traditional Dehnen and the Prugniel--Simien 
-- mostly because of its shallower central cusp.  

The full 5-parameter version of the Terzi\'c--Graham model should only be 
used to model galaxies with depleted cores, which means that it would 
{\it always} have shallower inner slopes than their Dehnen or Prugniel--Simien 
counterparts, and therefore be computationally very competitive with them, 
often even outperforming them (the exact performance depending on the level 
of depletion of the core).

The special case of the 5-parameter Terzi\'c--Graham, the 3-parameter 
Prugniel--Simien, should only be used to model galaxies without depleted 
cores.  In those cases, the computational price to pay over the traditional
Dehnen model is modest -- about $40-80\%$ over the meaninful range of
energies.  We believe that having a quantitative model appreciably more 
faithful to the physical problem is well worth the price. 

\section{Summary} \label{Section:Summary}

Motivated by the findings of earlier studies (Graham et al.\ 2002, 2003;
Terzi\'c \& Graham 2005) which demonstrate that the traditional power-law 
density models, such as 3-parameter Dehnen, are inadequate to describe 
density profiles whose slopes continuously vary as a function of 
radius -- as is the case for most elliptical galaxies and bulges -- we 
introduce a new family of realistic triaxial density--potential--force 
profiles for stellar systems and dark matter halos.  The new family is a 
5-parameter, two-regime model of inner power-law density core and the outer 
deprojected S\'ersic profile, with the boundary between the two regimes 
being the break radius $r_b$.  For extreme values of the break radius the 
model reduces to the 3-parameter deprojected S\'ersic $R^{1/n}$ profile 
($r_b \to \infty$) and the 2-parameter power-law core ($r_b \to 0$).  
The full 5-parameter model is invoked only when modeling elliptical galaxies 
with depleted cores; in other cases, which comprise the majority of observed 
elliptical galaxies and bulges, the 3-parameter triaxial deprojected S\'ersic
model suffices.  

In Section \ref{Section:Model}, we derived the potential and forces for the 
fully triaxial density model first presented in its spherical form by 
Terzi\'c \& Graham (2005).  Potential and forces are expressed in terms of 
easy-to-evaluate quadratures.  In Section \ref{Section:Code}, we developed 
an optimized suite of numerical routines designed for use in orbital 
integrations.  

The goal of this study was to develop and offer to the community an 
alternative to the traditional Dehnen model which is substantially more 
faithful to the physical problem (as measured by the quality of fits to 
deprojected light profiles) at an almost marginal computational expense.

\section{acknowledgments}

We are happy to thank Alister Graham for his helpful comments and suggestions.
We are grateful to Nicola Caon for providing us with the light--profiles 
for NGC 1379 and to Peter Erwin for kindly supplying us with the 
light--profile for NGC 3348.
B.J.S. is supported by NICADD grant G1A62214.

\appendix

\section{Spherical Limit} \label{ss:SphLimit} 

In the spherical limit $a=b=c=1$, $m=r$ and 
${\bar m}(\tau)={\bar r}(\tau)=r/\sqrt{\tau+1}$ and all quadratures above 
have analytic solutions given in terms of elementary and special functions:
\begin{eqnarray} \label{spec_fun_1sph_1}
H^I(x,y,z;1,1,1,q,s) & = & r^q \int\limits_s^{\infty} 
(\tau+1)^{-{3\over 2}-q} ~d\tau \nonumber \\
& = &{{2}\over{1+2q}} r^q (s+1)^{-{1\over 2}-q}, 
\end{eqnarray} 
\begin{eqnarray} \label{spec_fun_1sph_2}
H(x,y,z;1,1,1,q) & \equiv & H^I(x,y,z;1,1,1,q,0) \nonumber \\
& = & {{2}\over{1+2q}} r^q, 
\end{eqnarray} 
\begin{equation} \label{spec_fun_1sph_3}
R^I(1,1,1,s) = \int\limits_s^{\infty} (\tau+1)^{3 \over 2} ~d\tau =  
{2 \over{\sqrt{s+1}}},
\end{equation} 
\begin{equation} \label{spec_fun_1sph_4}
R(1,1,1) \equiv R^I(1,1,1,0) = 2,
\end{equation} 
\begin{equation} \label{spec_fun_1sph_5a}
T(x,y,z;1,1,1,s) = \int\limits_0^s 
{{\Gamma\left[n(2-p),b_n\left({{{\bar m}}\over{R_e}}\right)^{1/n}\right] d\tau}
\over{(\tau+1)^{3\over 2}}} \nonumber
\end{equation} 
\begin{eqnarray} \label{spec_fun_1sph_5b}
= & & 2 \Gamma\left[n(2-p),b_n\left({r\over R_e}\right)^{1/n}\right] \nonumber \\
& - & 2 {r_b \over r} \Gamma\left[n(2-p),b_n\left({r_b\over R_e}\right)^{1/n}\right] 
\nonumber \\
& + & 2 {R_e \over r} b_n^{-n}
\Gamma\left[n(3-p),b_n\left({r_b\over R_e}\right)^{1/n}\right] \nonumber \\
& - & 2 {R_e \over r} b_n^{-n}
\Gamma\left[n(3-p),b_n\left({r\over R_e}\right)^{1/n}\right],
\end{eqnarray} 
\begin{eqnarray} \label{spec_fun_1sph_6}
H_l^I(x,y,z;1,1,1,s) & = & 
\log r \int\limits_s^{\infty} 
(\tau+1)^{-{3\over 2}} ~d\tau 
\nonumber \\
& - & {1 \over 2} \int\limits_s^{\infty} {{\log(\tau+1)}\over
{(\tau+1)^{3\over 2}}} ~d\tau \nonumber \\
& = & {{2}\over{1+s}} \left[\log {{r}\over{\sqrt{1+s}}} -1\right], 
\end{eqnarray} 
\begin{eqnarray} \label{spec_fun_1sph_7}
H_l(x,y,z;1,1,1) & \equiv & H_l^I(x,y,z;1,1,1,0) \nonumber \\
& = & 2 \log r - 2, 
\end{eqnarray} 
\begin{eqnarray} \label{spec_fun_1sph_8}
H^I_2(x,y,z;1,1,1,1,q,s) & = & r^q \int\limits_s^{\infty} 
(\tau+1)^{-{5\over 2}-q} ~d\tau \nonumber \\
& = & {{2}\over{3+2q}} r^q (s+1)^{-{3\over 2}-q}, 
\end{eqnarray} 
\begin{eqnarray} \label{spec_fun_1sph_9}
H_2(x,y,z;1,1,1,1,q) & \equiv & H_2^I(x,y,z;1,1,1,1,q,0) \nonumber \\
& = & {{2}\over{3+2q}} r^q, 
\end{eqnarray} 
\begin{equation} \label{spec_fun_1sph_10}
S(x,y,z;1,1,1,1,s) = \int\limits_0^s 
{{\left({{\bar r}\over{R_e}}\right)^{-p} 
e^{-b_n\left({{{\bar r}}\over{R_e}}\right)^{1/n}}}
\over{(\tau+1)^{5\over 2}}} ~ d\tau \nonumber 
\end{equation} 
\begin{eqnarray} \label{spec_fun_1sph_11}
& = & \left(r\over R_e\right)^{-p} \int\limits_0^s  
{e^{-b_n\left({{{r}}\over{\sqrt{\tau+1} R_e}}\right)^{1/n}}
\over{(\tau+1)^{{{5-p}\over 2}}}} ~ d\tau \nonumber \\
& = & 
2 n b_n^{n(p-3)} \left({r \over R_e}\right)^{-3} 
\Gamma\left[n(3-p),b_n\left({r\over {R_e\sqrt{s+1}}}\right)^{1/n}\right] 
\nonumber \\
& - & 
2 n b_n^{n(p-3)} \left({r \over R_e}\right)^{-3} 
\Gamma\left[n(3-p),b_n\left({r\over {R_e}}\right)^{1/n}\right].
\\ \nonumber
\end{eqnarray} 
The result for $T(x,y,z;1,1,1,s)$ is derived after integration by parts.
$\tau_b$ is easily found from $r_b=r/\sqrt{\tau+1}$.

After some straightforward yet tedious algebra, the equations for mass 
density, potential and forces reduce to those of Terzi\'c \& Graham (2005), 
their equations (6), (8) and (14), respectively.

\section{Initial Conditions Generator} \label{sec:ICs}

\begin{figure*}
\begin{center}
\includegraphics[width=6in]{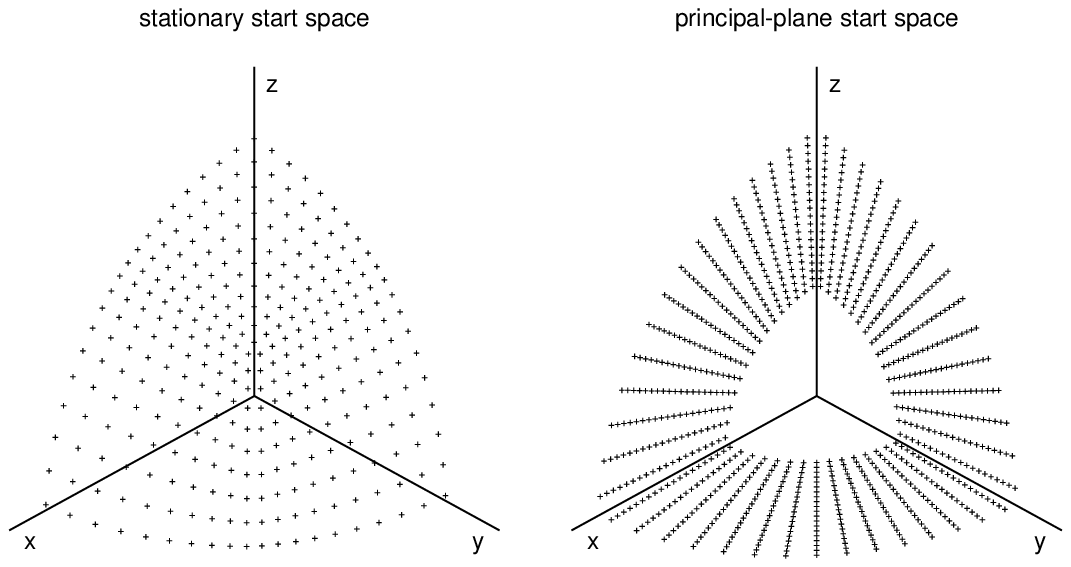}
\vskip-40pt
\caption{Stationary and principal-plane start spaces.}
\label{figss}
\end{center}
\end{figure*}

Sampling of the initial condition space is done by employing a two-fold
two-dimensional start space, following Schwarzschild (1993):
the {\it stationary start space} which contains initial conditions starting
from equipotential surfaces with zero velocities; and 
the {\it principal-plane start space} which consists of radially stratified
initial conditions which pierce one of the three principal planes with the
velocity vector normal to the plane.  These start spaces are designed to
sample different types of orbits arising in triaxial potentials: 
stationary start space picks up orbits which have zero-velocity turning 
points, such as boxes and resonant boxlets, while the principal-plane 
start space selects mostly tube orbits \cite{S93,T02}.  
The two start spaces are shown in Fig.~\ref{figss}.

The spherical angles of the stationary start space locations are chosen as 
in Schwarzschild (1993), with the addition that the number of points can be 
chosen using a parameter ${\tt stat\_N}$, so that the total number of points 
on this equipotential surface will be ${\tt 3 \ stat\_N^2}$.  The 
zero-velocity radius of this potential in the given direction is then found, 
resulting in an exact coordinate for the given initial condition.

We calculated zero-velocity radii using a secant-method search to find the 
root of $\Phi(r,\theta,\phi)-E=0$, given the direction angles $\theta$ and 
$\phi$.  This method results in very fast convergence to the solution, even 
on the unbounded interval $0 < r < \infty$.  Proper convergence of this 
method does assume, however, that the $\Phi(r,\theta,\phi)$ function is 
monotonically increasing with radius $r$, which is a condition that is met 
by the three potential functions considered in this study.

Initial conditions in the principal-plane start space are calculated in a 
similar manner to Schwarzschild (1993) as well.  The parameters 
${\tt plane\_N\_theta}$ and ${\tt plane\_N\_r}$ determine the number of points 
sampled.  First, the angle $\theta$ is subdivided into ${\tt plane\_N\_theta}$ 
regions from 0 to $\pi/2$.  The centers of each of these regions are chosen as 
direction vectors within the plane for choosing initial conditions.  Along each 
direction vector, the zero-velocity radius ${\tt rmax}$ is computed as 
described above.  Next, an inner cutoff radius is determined by using a 
fraction ${\tt plane\_frac\_rmin}$ such that 
${\tt rmin=plane\_frac\_rmin \times rmax}$.  
This inner cutoff fraction is a parameter that can be chosen from $0$ to $1$ in 
order to reduce the number of orbits that are duplicated at small radii.  The 
initial conditions are then chosen by subdividing the region ${\tt rmin}$ to 
${\tt rmax}$ into ${\tt plane\_N\_r}$ sections, and selecting the center of 
each section as the location of the initial condition.  The velocity is then 
chosen normal to the plane such that the total energy is equal to the desired 
energy level.  This is done for each of the 3 principal planes, resulting in 
${\tt 3 \times plane\_N\_theta \times plane\_N\_r}$ initial conditions 
generated in this start space.

\section{Dehnen Model} \label{apdx:Dehnen}
Dehnen model \cite{D93,Tetal94} is a double-power-law, $3$-parameter 
density model, defined by its steepness of the inner cusp $\gamma_D$, break 
radius $r_a$ which marks the transition between the two power-law regimes, 
and the density at the break radius $\rho(r_a)$:
\begin{equation} \label{Dehnen_rho}
\rho (m) = 2^{4-\gamma_D} \rho(r_a)
\left({m\over r_a}\right)^{-\gamma_D} \left(r_a+m\right)^{\gamma_D-4}.
\end{equation}
The total mass is easily found to be
\begin{equation} \label{Mtot_d}
M_D={{4\pi a b c}\over{3-\gamma_D}} r_a^3 2^{4-\gamma_D} \rho(r_a).
\end{equation}
The corresponding potential is given by 
\begin{equation} \label{Dehnen_phi}
\Phi (x,y,z) = -{{G M_D}\over{2(2-\gamma_D)r_a}} 
\int\limits_0^{\infty} 
{{P({\bar m})} \over{d_1(\tau;a^2,b^2,c^2)}} ~ d\tau,
\end{equation}
where
\begin{eqnarray} \label{pfunc}
P({\bar m}) \equiv 1 &-&(3-\gamma_D)\left({{\bar m}\over{r_a+{\bar m}}}\right)^{2-\gamma_D} \nonumber \\
&+&(2-\gamma_D)\left({{\bar m}\over{r_a+{\bar m}}}\right)^{3-\gamma_D}
\end{eqnarray}
This equation is, again, transformed into a proper integral via transformation
$\xi=(\tau+1)^{-1/2}$, to obtain
\begin{equation} \label{Dehnen_phi2}
\Phi (x,y,z) = -{{G M_D}\over{(2-\gamma_D)r_a}} 
\int\limits_0^1
{{P({\bar m})} \over{{\tilde d}_1(\xi;a^2,b^2,c^2)}} ~ d\xi.
\end{equation}
Similarly, the forces are found to be, after the change of variables, 
\begin{eqnarray} \label{Dehnen_f}
F_{q_i} (x,y,z) & = & -G M_D (3-\gamma_D) r_a q_i \nonumber \\
& \times & \int\limits_0^1
{{{{{\bar m}^{-\gamma_D}}{(r_a+{\bar m})^{\gamma_D-4}}} \xi^2}
\over
{{\tilde d}_2(\xi;k_i^2;a^2,b^2,c^2)}} ~ d\xi,
\end{eqnarray}
where $k_1=a$, $k_2=b$, $k_3=c$ and $q_1=x$, $q_2=y$, $q_3=z$.

\label{lastpage}
\end{document}